\documentclass[12pt]{JHEP3} 
\usepackage{amsfonts}
\usepackage{amssymb}
\usepackage{cite}
\def\be{\begin{equation}} \def\ee{\end{equation}}
\def\bea{\begin{eqnarray}} \def\eea{\end{eqnarray}}
\def\ba{\begin{array}}
\def\ea{\end{array}} \def\ben{\begin{enumerate}}
\def\een{\end{enumerate}}
  
\def\lll{\label}

 \title{ Duality symmetric massive type II theories in $D=8$ and $D=6$
\thanks{Work 
supported by:
AvH -- the Alexander von Humboldt foundation.}}

\author{
 Harvendra Singh\\
 Fachbereich Physik, Martin-Luther-Universit\"at Halle-Wittenberg,\\
Friedemann-Bach-Platz 6, D-06099 Halle, Germany}

\abstract{
We study $T^2$  compactification
of  massive type IIA supergravity  in presence of  
possible Ramond-Ramond (RR) background fluxes. 
The resulting theory in $D=8$ 
is shown to possess full $SL(2,R)\times SL(2,R)$ T-duality symmetry 
similar to the massless case.
It is shown  that elements of duality symmetry interpolate between 
massive type IIA compactified on $T^2$ and ordinary type IIA
compactified
on $T^2$ with RR 2-form flux.
We also discuss relationship between M-theory vacua and massive 
type IIA 
vacua. The D8-brane is found to correspond to M-theory `pure gravity'
solution which is a direct
product of 7-dimensional  Minkowski space and a 4-dimensional 
 instanton.  We also construct D6-D8 bound state which preserves $1/2$
 supersymmetries. 
We then discuss massive IIA  compactification on $T^4$ and point
 out that when all possible RR fluxes on $T^4$ are turned on the
 six-dimensional theory appears to
 assume a nice $SO(4,4)$ invariant form.}

\preprint{hep-th/0109147}
\keywords{ strings, supergravity, compactification, dualities}

\begin{document}
\newcommand{\eqn}[1]{(\ref{#1})}
\def\cC{{\cal C}} 
\def\cG{{\cal G}} 
\def\cd{{\cal D}} 
\def\a{\alpha}
\def\b{\beta}
\def\g{\gamma}\def\G{\Gamma}
\def\d{\delta}\def\D{\Delta}
\def\ep{\epsilon}
\def\e{\eta}
\def\z{\zeta}
\def\t{\theta}\def\T{\Theta}
\def\l{\lambda}\def\L{\Lambda}
\def\m{\mu}
\def\f{\phi}\def\F{\Phi}
\def\n{\nu}
\def\p{\psi}\def\P{\Psi}
\def\r{\rho}
\def\s{\sigma}\def\S{\Sigma}
\def\ta{\tau}
\def\x{\chi}
\def\o{\omega}\def\O{\Omega}
\def\k{\kappa}
\def\pa {\partial}
\def\ov{\over}
\def\br{\nonumber\\}
\def\ud{\underline}

\section{Introduction}

Recently considerable attention has been focused on the
 study of gauged/massive supergravity theories owing
to their importance in  AdS/CFT analysis \cite{maldacena}. 
In traditional methods gauged supergravity theories can be constructed
out of 
their ungauged versions either by gauging  
 a subgroup of the $R$-symmetry group 
and/or by gauging the isometries of the scalar manifold
 using the vector fields in the 
spectrum \cite{gsugra}. 
This procedure is incorporated in a manner such that it 
does not change the overall particle spectrum
and the number of supersymmetries in the theory.  
However it does, generically,
 change the properties of the ground state.
In most of the cases, if a Minkowskian spacetime is a 
 solution  of the ungauged theory,  it ceases
to be a ground state in the case of gauged supergravity.
Instead the supersymmetric ground states are
often of the anti-de-Sitter (AdS)   
or domain-wall type solutions. 
Gauged supergravities with AdS-ground states
have been focus of much attention in AdS/CFT correspondence as they can
be
derived through a compactification of ten or eleven dimensional
 supergravity theories on spacetimes involving AdS subspaces, e.g.,
 $AdS_{p+2}\times S^{D-2-p}$ which are near horizon geometries of
$p$-branes. 

We are interested here in the study of massive
supergravities which are closed relatives
of gauged supergravities. 
In these theories some of the vector 
(or tensor) fields become massive upon eating other fields in 
their massless spectrum, analogous to
 a  Higgs mechanism. In this procedure again the total degrees of
 freedom  remain unaltered and so does the number of supercharges. 
A well studied  example of a massive theory  
is the massive type IIA supergravity (m-IIA)
in $D=10$ constructed by Romans \cite{roma}.
In string theory massive supergravities
typically can arise in lower dimensions through generalized
Scherk-Schwarz
reduction \cite{ss},
provided that some field strength $dA_{(p)}$
of the $p$-form tensor field $A_{(p)}$ is given a 
non-trivial background value (flux) along the
compact directions \cite{berg1}. 
Such background fluxes can be turned on
consistently if the action and the field equations
depend on $A_{(p)}$ 
only through its field strength $dA_{(p)}$. 

Our purpose here is to study massive supergravities 
in the context of string dualities and D$p$-branes. 
The massive type IIA supergravity has a domain wall solution
which  preserves $1/2$ of the
32 supercharges \cite{berg1} and has been given an
interpretation of a type IIA D-8-brane.
This observation has led to the search for possible duality connections
involving massive supergravity theories analogous to the existing
duality symmetries in the ordinary (massless) cases.  
This required the construction of new massive 
supergravities in lower dimensions through generalized dimensional
reduction 
\cite{berg1,cow,blo,bs,mo,lavpop1,lavpop,bre, lupop, kkm,km,cvetic,hull, 
hls, martin,martin1,janssen}. 
However, it has still remained an interesting open question, 
  to what extent the generalized compactifications do
 respect the duality properties of the massless cases.
In a parallel line of developments
Calabi-Yau compactifications with background fluxes 
have also been studied 
because of their
phenomenological properties 
\cite{DSHD,S,PS,BB,JM,LOW,GVW,DRS,G,TV,M2,ST,GSS,HLM,
CKLT,EW,LOW2,LOSW,NOY,BG,CK}. 
One finds that background
fluxes typically generate a potential 
for some of the moduli fields of the 
theory without fluxes and as a consequence the moduli space
-- and hence the arbitrariness of the theory --
is reduced. In addition the resulting ground states 
can break supersymmetry spontaneously.

In this paper we
study a generalized $T^2$ and $T^4$ reduction
of ten-dimensional massive type IIA supergravity
with all possible R-R background fluxes turned on.
Our goal is to investigate the fate of the
perturbative $O(d,d)$ 
duality symmetries and the non-perturbative
S-duality.
We  also discuss the relation of massive type
IIA theory  with M-theory.
The paper is organized as follows. 
In section~2 we briefly recall  massive type IIA sugra
and fix our conventions to be used for the cases of toroidal reductions. 
Section~3 covers the compactification of massive type IIA theory on
$T^2$.
We find that provided RR 2-form flux is 
turned on the resulting eight-dimensional
theory can be presented in a manifestly $SL(2,R)\times SL(2,R)$
invariant 
form. In section~4 we study the vacuum solutions
of this massive 8-dimensional supergravity and 
relate them to the solutions of ordinary IIA
by using the elements of T-duality group. We  uplift these ordinary IIA
vacua to eleven dimensions. Particularly the 
D8-brane is shown to correspond to
a pure gravity solution in eleven dimensions which contains an 
domain-wall-type instanton line element. 
Thus perturbative T-duality symmetry
interpolates 
between vacua of massive and massless type IIA theories compactified 
on $T^2$. 
This property also allows us to further relate massive type II vacua
to eleven dimensional solutions.
In section~5
we obtain a D6-D8-brane bound state which preserves 16 supercharges. 
In section~6 we outline the compactification on $T^4$ and point out that
the 
 massive IIA
theory on $T^4$ will have a full $SO(4,4)$ symmetry provided all the RR
fluxes
for all the fields are turned on.
Finally we have summarized our results in the section~7.

\section{Review}
The type IIA supergravity in ten dimensions, which describes the low
energy limit of type IIA superstrings,
contains in the massless bosonic spectrum the graviton
$\hat g_{MN}$, the dilaton
$\hat\f$,  NS-NS two-form $\hat B_{(2)}$, a R-R one-form $\hat A_{(1)}$ 
and a R-R three-form $\hat C_{(3)}$. The fermionic
fields consist of two gravitini and two Majorana ${1\ov2}$-spinors.  
Massive type IIA supergravity (m-IIA) \cite{roma} is a
generalization of that to include a mass term for the $\hat B$-field
without 
disturbing the supersymmetry. More precisely, 
the $\hat B$-field becomes massive through a
Higgs type mechanism in which it eats the vector field $\hat A$. 
The supersymmetric bosonic action for massive IIA theory in the string
frame can be written as \footnote{ Our conventions are same as in
  \cite{hls}  where  
 every product of forms is understood 
 to be a wedge product. We denote a $p$-form  with a lower index like
$(p)$ 
and later drop it, for simplification, when it becomes obvious.}
\begin{eqnarray} \label{massive2a}
S&=&\int \bigg[ e^{-2\hat\f}\left\{{1\ov 4}\hat
R~^{\hat\ast}1+d\hat\f~^{\hat\ast} d\hat\f -
{1\over2}\hat H_{(3)} ~^{\hat\ast}\hat H_{(3)}\right\} 
-{1\over2}\hat F_{(2)} ~^{\hat\ast}\hat F_{(2)} -{1\over2}\hat F_{(4)}
~^{\hat\ast}\hat F_{(4)}- {m^2\ov 2 }~^{\hat\ast} 1 \br
&& + d\hat C_{(3)} d\hat C_{(3)} \hat B_{(2)} + 2d\hat
C_{(3)} d\hat A_{(1)} \hat B_{(2)}^2+{4\ov3}d\hat A_{(1)} d\hat A_{(1)}
\hat B_{(2)}^3
+{4\ov3}m d\hat C_{(3)}\hat B_{(2)}^3 \br 
&&+2m d\hat A_{(1)} \hat B_{(2)}^4+ {4\ov5}m^2\hat B_{(2)}^5\bigg]\ ,
\lll{1a}
\end{eqnarray}
where $m$ is the mass parameter. The 
 various field strengths in the action \eqn{1a}
are given by
\be
\hat H_{(3)}=d\hat B_{(2)}\ ,
\qquad 
\hat F_{(2)}= d\hat A_{(1)} +2m\hat B_{(2)}\ ,
\qquad
\hat F_{(4)}= d\hat C_{(3)}
+2\hat B_{(2)}d\hat A_{(1)}+2m\hat B_{(2)}^2\ .
\lll{a22}\ee
Note that potentials
$\hat A$ and $\hat C$  appear only through their derivatives 
in the action \eqn{1a} and thus obey the standard 
$p$-form gauge invariance $A_{(p)} \to A_{(p)} + d\lambda_{(p-1)}$.
The two-form $ \hat B$ on the other hand also appears
without derivatives but nevertheless the `Stueckelberg'
gauge transformation
\be\label{stueck}
\delta \hat A=-2m\lambda_{(1)}\ ,\qquad
\delta \hat B=d\lambda_{(1)}\ ,\qquad 
\delta\hat C=-2\lambda_{(1)} d\hat A\ 
\ee
leaves the action invariant. 

As we shall turn to the compactification in the later sections let
us  recall here some facts about toroidal compactifications. 
Standard Kaluza-Klein 
reduction considers the theory in a spacetime background 
$M_D\times T^d$, where $M_D$
is a non-compact $D$-dimensional manifold with 
Lorentzian signature
while $T^d$ is a $d$-dimensional compact torus.
This ansatz is consistent whenever the spacetime
background satisfies the $(D+d)$-dimensional field equations.
However, for massive type IIA there are no direct product $M_D\times
T^d$ solutions; instead the ground states are domain-wall (D-8-brane)
solutions
\cite{berg1}, see section 4. 
It has been discussed in \cite{berg1} that Kaluza-Klein
reduction can still be carried out for such theories.  
The $T^1$ compactification of massive type IIA has been discussed 
at great length in \cite{berg1}. These ideas have also 
 been applied to K3 compactifications recently \cite{hls}.
 
Thus for the 
10-dimensional zehnbein we  take the standard toroidal ansatz
\be\label{metric}
\hat e^{\hat a}_{M}\, (x,y)= \left( \ba{cc}e^a_\m(x)&~~~e^i_m K^m_\m\\
0&e^i_{m}\ea\right)\ ,
\ee 
where coordinates $y^m~(m=1,...,d)$ are tangent to the tori. 
The internal metric on tori
is  given by $G_{mn}(x)= e_m^i \d_{ij}e_n^j$ while
the D-dimensional Minkowski metric is $ g_{\m\n}(x)= e_\m^a
\eta_{ab}e_\n^b$. 
$K^m_\m$ are the
Kaluza-Klein gauge fields. Let us 
define  gauge invariant $d$ one-forms  $\eta^m_{(1)}=dy^m+K^m_{(1)}$.

The standard toroidal ansatz for
 the dilaton and the NS-NS two-form $\hat B$ is 
\be
\hat\f(x,y)=\hat \f(x)\ ,\qquad
\hat B_{(2)}(x,y)=B_{(2)}(x)+ \bar A_m(x)~ \e^m+{1\over 2}B_{mn}(x)~\e^m 
 \e^n\ ,
\ee
where $B_{(2)}=\bar B_{(2)}-{1\ov2}\bar A_m K^m$ with $\bar B$  being 
a 2-form in $D$ 
dimensions, $\bar A_m$ are $d$ vector
potentials      
and the $B_{mn}$ represents scalar fields antisymmetric in $m,n$ 
indices ($m,n=1,...,d$). With these ans\"atze the NS-NS part of the
action
 \eqn{massive2a} reduces
as follows \cite{ms}
\bea
\int  
 && e^{-2\f}\bigg[{1\ov 4}
R~^{\ast}1+d\f~^{\ast} d\f -
{1\over2} H_3 ~^{\ast} H_3 
+{1\ov16}dG^{mn}~^\ast dG_{mn}-{1\ov4}dB_{mn}~^\ast dB_{pq}G^{mp}G^{nq}  
\br &&-{1\ov8}dK^m~^\ast dK^nG_{mn}  
-{1\ov 2}(d\bar A_m-B_{mp}dK^p)~^\ast (d\bar A_n-B_{nq}dK^q)G^{mn}
\bigg]~,
\eea
with
\be
\label{deff1}
2\f=2\hat\f-{1\ov2} \ln~ G  ,\qquad H_{(3)}=d\bar B -{1\ov2}
(\bar A_m dK^m+K^md\bar A_m)\ , \ee
where $G$ represents the determinant of the internal metric.

In the next sections we shall consider the 
specific cases where we consider the compactifications of m-IIA on
even tori. We are focusing on even tori as we are interested in
generalized reduction with the presence of RR-fluxes. Note that type IIA
involves only even-form field strengths in the RR sector.  
The compactification of m-IIA on $T^1$ has been discussed in 
\cite{berg1} which also involves
a generalized reduction of type IIB in order to study duality symmetry. 
Similarly a $T^3$ reduction of m-IIA has to
deal with the generalized reduction of type IIB on $T^3$.      

\section{Compactification on $T^2$}

Here we  specifically consider the case of compactification on a
2-torus.
For the one-form $\hat A_{(1)}$ and the three-form $\hat
C_{(3)}$ we would consider  a generalized 
Kaluza-Klein ansatz  where background fluxes are considered.
This generalization is possible since  
$\hat A_{(1)}$ and $\hat
C_{(3)}$  appear in the action \eqn{1a} 
only through  derivatives, therefore an appropriate background value
can be consistently turned on. 
We take the following  generalized ansatz
\bea
&& \hat A(x,y)=  A_{(1)}(x)+  A_{(0)m}(x,y) dy^m  ,\br
&&\hat C(x,y)= C_{(3)}(x)+ 
T_{(2)m}(x) dy^m+{1\ov2} V_{(1)mn}(x) dy^mdy^n.
\lll{ansatz1}
\eea
Note that the scalar-forms are allowed to retain a  
dependence on the coordinates of the torus unlike in standard toroidal 
reduction.  The consistency of toroidal reduction requires it to be at
most
a linear dependence on the torus coordinates, 
we fix it to be
 \be
A_m(x,y)=a_m(x)-{1\ov2}m_{mn} y^n,
\lll{ansatzz}\ee
 where constants $m_{mn}$ are
antisymmetric in indices ($m,n=1,2$). This will give us
\bea
&& \hat d\hat A= \cd A + \cd a_m \eta^m +{1\ov 2}m_{mn} \e^m\e^n,\br
&&\hat d\hat C=\cd C_{(3)}(x)+\cd T_{(2)m} \eta^m +{1\ov2} 
d V_{(1)mn}\e^m\eta^n 
\lll{ansatz1a}
\eea
where various $\cd$-derivatives are defined as 
\bea
&& \cd A= dA -da_m K^m +{1\ov2}m_{mn}K^mK^n~,~~~ \cd a_m=da_m + m_{mn}
K^n, \br
&& \cd T_{m }= dT_m +d V_{mn} K^n~,~~~
 \cd C= dC -dT_m K^m+{1\ov 2} dV_{mn}K^m K^n.
\eea
Note that various forms can be distinguished by their symbols and 
the internal indices they carry.
Thus we see explicitely that even if the potential $\hat A$ in
\eqn{ansatz1}
depends upon the  torus 
coordinates, the derivatives in \eqn{ansatz1a} do not. 
This dependence  also drops out in the action and 
 the field equations.  
Through above  generalized ansatz we have effectively 
introduced  one new  parameter $m_{12}$ in 
the guise of 2-form flux. 
This generalization has been possible only because  
$\hat A_1$ appears in the action 
 covered with derivative and $T^2$ has one 2-cycle along which
 an appropriate background flux
could be turned on. 

The massive field strengths  in \eqn{a22} become
\bea
&&\hat F_{(2)}=(\cd A+2m B) +(\cd a_n+2m\bar A_m)\e^m +{1\over2} 
(m_{mn}+2mB_{mn})\e^m\e^n~, \br 
&&\hat F_{(4)}=(\cd C+ 2B\cd A+2m BB) +(\cd T_n+ 2B\cd a_m+ 
2 \bar A_m\cd A+4mB 
\bar A_m )\e^m 
\br &&\hskip1.5cm +{1\over2} 
(\cd V_{mn} -4\bar A_m\cd a_n +2B_{mn}\cd A +2Bm_{mn} 
+4 mBB_{mn}-4m\bar A_m\bar A_n)\e^m\e^n~. \br 
\eea

Altogether,
the bosonic spectrum of the reduced 8-dimensional
theory consists of the graviton $g_{\m\n}$, dilaton $\phi$, 2-form
$\bar B$, 4 scalars, and 4 vectors $(\bar A_m,~K^m)$ in the NS-NS
sector. From the R-R sector we have 2 scalars $a_m$, 2 vectors
$(V_{12},A)$, 
2 tensors
$T_m$ and one 3-form $C$ whose field strength is (anti)self-dual in
$D=8$. 
Also we have two parameters $m_{12}$ and $m$.
This is the bosonic
content of maximal type II supergravity theory in $D=8$.
In the massless case these fields fit into various representation of the
T-duality group $SL(2,R)\times SL(2,R)\sim SO(2,2)$.
Here too various fields combine into the   
$SL(2,R)\times SL(2,R)$ representations  as
\bea
&&
A^{ru}_{(1)}=(A^{r1},A^{r2}),~~A^{ru=1}=(\bar 
A_1,~\bar A_2),~~A^{ru=2}=({K^2\ov2},~~-{K^1\ov2}),\br
&&
a^r_{(0)}=(a_1,a_2),~~t^r_{(2)}=(T_1,~T_2),~~
{\cal A}^u_{(1)}=(-V_{12}, A),~~m^u=(-m_{12},~m),
\eea
where indices $r=1,2$ belong to first $SL(2,R)$ 
while indices $u=1,2$ belong to the second $SL(2,R)$ group.
Note that the mass and flux parameters also fit into a 
fundamental representation of $SL(2,R)$.

In order to obtain the action of the massless
modes for this theory we
substitute the ansatz \eqn{metric}-\eqn{ansatz1a} into the action
\eqn{1a}. 
The resulting eight-dimensional bosonic action reads in the kinetic part
 
\bea
S_{D=8}&=&\int\bigg[{1\ov4}e^{-2\f}\bigg\{ R~^\ast 1+4~ d\f~ ^\ast d\f -
2~ H_3 ~^\ast H_3-2~ dA^{ru}~^\ast dA^{sv}
M^{-1}_{rs}{\cal M}^{-1}_{uv} \br 
&&
+{1\over 4} {\rm Tr}~ d{\cal M}^{-1} ~^\ast
d{\cal M}+ {1\over 4} {\rm Tr}~ d{ M}^{-1} ~^\ast
d{ M})\bigg\}
-{1\over2} \cG ~^\ast \cG ~\sqrt{G}\br &&
-{1\over2} {\cal F}_{(3)}^r ({ M}^{-1})_{rs}~^\ast 
{\cal F}_{(3)}^s 
-{1\over2} {\cal F}_{(2)}^u ({\cal M}^{-1})_{uv}~^\ast 
{\cal F}_{(2)}^v 
\br &&
-{1\over2} {\cal F}_{(1)}^r ({ M}^{-1})_{rs}~^\ast 
{\cal F}_{(1)}^s 
-{1\over2} m^u({\cal M}^{-1})_{uv}~^\ast m^v \bigg]+S_{C-S}~,
\lll{action8}
\eea
where the Chern-Simon part of the action is
\bea
&&S_{C-S}= \br
&&\int
- b \cG \cG +\cG \left\{ -2(\cd T_m +2\cd a_m B+\cd A 
\bar A_m+2mB\bar A_m)\bar A_n+dV_{mn}B+m_{mn}BB \right\}\ep^{mn}\br
&&~~  +\bigg[-\cd T_m\cd T_nB-\left\{ 2\cd T_m\cd a_n+\cd V_{mn}\cd A 
-2\cd A\cd A B_{mn}\right\}(B)^2 \br &&~~
+\left\{ 4 m \cd A B_{mn}-{4\ov3}m\cd V_{mn}-{2\ov3}m_{mn}\cd 
A-{4\ov3}\cd a_m\cd a_{n}\right\}(B)^3\br
&&~~+(2m^2B_{mn} -m~ m_{mn})(B)^4\bigg]\ep^{mn}~,
\lll{action88}
\eea
with $\ep^{12}=1$ and we have defined $b\equiv -B_{12}$, 
$$\label{deff}
2\f=2\hat\f-{1\over 2} \ln~ G   ,\qquad H_{(3)}=d\bar B + 
A^{ru}dA^{sv}L_{rs}{\cal L}_{uv}\ . $$ 
Various R-R field strengths in the above action are 
\bea
&&{\cal F}_{(1)}^r=da^r+2m^u~A_u^r\br
&&{\cal F}_{(2)}^u=d{\cal A}^u -2 da^r A_r^u- 2m^vA_v^rA_r^u +2 m^u\bar
B,\br
&&{\cal F}_{(3)}^r=dt^r+2d{\cal A}^u A_u^r-2da^sA_s^uA_u^r -{4 \ov3} 
m^vA_v^sA_s^uA_u^r+ 2\bar B {\cal F}_{(1)}^r\br
&& \cG_{(4)}=\cd C+2B\cd A+2m BB .
\lll{action88a}
\eea
The indices $r$ and $u$ can be raised or lowered by the use of 
 two $SL(2,R)$ metrics $L$ and ${\cal 
L}$, respectively. The two metrics are given by 
$$ L_{rs}=\left(\ba{cc} 0 &-1\\ 1 &0\ea\right)\equiv {\cal L}_{uv}\ . $$
The uni-modular matrices which belong to two  
$SL(2,R)/SO(2)$  
cosets are given by
\be
M^{-1}=\sqrt{G}\left(\ba{cc} G^{11} &~~G^{12}\\ G^{21} 
&~~G^{22}\ea\right),~~~~{\cal M}^{-1}= 
{1\over \sqrt{G}}\left(\ba{cc}  
1 &~~~2b\\ 2b &~~~4(b)^2+G\ea\right),
\ee
and they satisfy  $M^T L M=L,~~{\cal M}^T {\cal L}{\cal 
M}={\cal 
L}~.$
 
Under an $SL(2,R)$ transformation
\bea
&&{\cal M}\to \L{\cal M}\L^T, {\cal A}^u\to \L^u_{~v}{\cal A}^v, 
{ A}^{ru}\to \L^u_{~v}{ A}^{rv}~,\br && m^u\to\L^u_{~v}m^v~,
\lll{trans}
\eea 
with $ \L~{\cal L}~\L^T={\cal L}$. The
same is true for other $SL(2,R)$ group with acts upon $r,s$ indices.

Note that the kinetic terms in the action except the terms involving 
4-form field strength ${\cal G}$ 
remain invariant under the action of above T-duality group. We 
have not provided the explicit invariant form for the Chern-Simon terms
which
also remain invariant provided 
the flux $m_{12}$ and mass $m$ transform as an $SL(2,R)$ 
doublet.  It remains to be seen if the field equations 
and the Bianchi identity for 3-form 
potential $C$   transform covariantly.
Let us write down the field equation for $C$. From the 
8-dimensional action in \eqn{action8}  we get  
\bea
&&d\bigg[-(\sqrt{G}~^\ast \cG +2b~ \cG) + \big[-2(\cd T_m +2\cd a_m
B+\cd 
A \bar A_m+2mB\bar A_m)\bar A_n\br &&\hskip.5cm
+ dV_{mn}B+m_{mn}BB\big]\ep^{mn}\bigg]=0\equiv 
d(d\tilde C_{(3)})~,
\lll{a2a}
\eea
where $\tilde C_{(3)}$ is defined to be the dual 3-form potential.
If we now define  $ \cG^1= -\sqrt{G}~^\ast\cG -2b~ \cG$ 
and $\cG^2 = \cG$, then 
field equation \eqn{a2a} simply becomes a Bianchi identity for $\tilde
C_{(3)}$. 
This  Bianchi for
$\tilde C_{(3)}$
 and the Bianchi identity for $ C_{(3)}$, which can be derived from its
field 
strength in \eqn{action88a},  
form an $SL(2,R)$ covariant set of equations. From this 
we can  write down $SL(2,R)$ 
4-form field strength as
\bea
\cG^u&=& d\cC^u -2 dt_r A^{ru} -2 d{\cal A}_vA^{v}_rA^{ru}+{4\ov 
3}da_rA^r_vA^v_sA^{su} \br
&&~+{2\ov3} m_wA^w_rA^r_vA^v_sA^{su} + 2 {\cal F}_{(2)}^u \bar B + 2m^u
(\bar 
B)^2~,
\eea
where 3-forms are $ \cC^u=(\tilde C, C)$ which transform as $SL(2,R)$ 
vector.

The action  \eqn{action8}  possesses Stueckelberg gauge invariances 
which 
is obvious from the investigation of the field strengths in 
eq.\eqn{action88a}. Through these gauge invariaces, the vector fields 
$A^{ru}$ can eat the scalars $a^r$ and can become massive. Similarly the 
tensor field $\bar B$ can eat one of the vector fields ${\cal A}^u$ and 
can become 
massive. Explicit gauge transformations of the fields can be derived
from 
those in \eqn{stueck}.   

This completes our analysis of the eight-dimensional massive type II 
supergravity action which we have 
shown to possess an explicit $SL(2,R)\times SL(2,R)$ T-duality symmetry 
provided 
the flux $m_{12}$ and the m-IIA mass $m$ transform as $SL(2,R)$ 
doublet. Note that this restoration of the  T-duality symmetry of the
massive II theory in $D=8$ is direct consequence of our ansatz in
\eqn{ansatzz}. Under this structure the mass and the 
flux behave in identical manner.
 If we start without any flux we can generate a flux by 
making an $SL(2,R)$ rotation, see next section. 
In other words, if we compactify m-IIA  on $T^2$ without RR flux, the 
resulting massive theory in $D=8$ can be mapped to ordinary IIA theory 
compactified on $T^2$ with RR 2-form flux, by making use of the duality 
symmetry.

We are not  surprised with this identification between flux and 
mass. It is the repetition of the story when m-IIA is compactified 
on a circle of radius $R$ and type IIB strings compactified on a circle
of
radius $1/R$ along with RR 1-form (axionic scalar) flux. This 
identification has been crucial in order to make massive T-duality 
 in $D=9$ manifest \cite{berg1}.
Similar phenomenon has  been shown to be repeated for the case of 
compactification on $K3$ manifold also \cite{hls}.

\section{D8-brane vs M-theory instanton}

The ten-dimensional massive IIA supergravity theory has  D8-brane
(domain-wall) solutions which preserve sixteen supercharges
\cite{berg1}. 
In the string frame  metric the solution is given by
\bea
&&d\hat s^2= H^{-1/2} (-dt^2 + dx_1^2+\cdots +dx_6^2+
dy_1^2+dy_2^2  
) +H^{1/2}dz^2,
\br &&2\hat\f=-{5\ov2}\ln H,
\lll{wall}
\eea
where 
$H=const. +{2m}|z-z_0|$ is a harmonic function of only the transverse
coordinate $z$ and all other fields have vanishing background values,
$z_0$
refers to the location of the domain-wall.
We  compactify this solution  by 
wrapping two of its world-volume directions, say $y_1, y_2$,  on $T^2$.  
The corresponding  8-dimensional domain-wall (or 6-brane) solution 
can be written down using our analysis  of the last section
\bea\label{sixs}
&&ds_8^2= H^{-1/2} (-dt^2 + \sum_{i=1}^6dx_i^2) +H^{1/2}dz^2\ ,
\br &&2\f= 2\hat\f-{1\over2} \ln G=-{2} \ln H \ ,
\br &&m^u=(0,~m)\ ,~~~{\cal M}={\rm diag}(H^{-1\ov2}, H^{1\ov2})\ , 
\eea
while  other 8-dimensional background fields
vanish. 
Clearly this vacuum configuration corresponds to the situation when 
there is no background flux.
The solution (\ref{sixs}) still has $16$ unbroken supersymmetries. Let
us note
that similar domain-wall type solutions appear in various situations and 
in one such case  domain-wall solution  are discussed in 
\cite{cow} where a massive 8-dimensional theory is obtained 
from generalized reduction from 9-dimensional type II theory to eight 
dimensions. 

Now, by applying an $SL(2,R)$ 
transformation \eqn{trans} on the background fields in \eqn{sixs}
new solutions with non-trivial R-R flux can be generated.
Let us consider the special case
of $SL(2,R)$ transformation given by 
\be\label{U3}
\L=\left(\ba{cc} 0&-1\\  1&0 \ea\right)\ .
\ee
Inserting $\L$ and the configuration \eqn{sixs} in \eqn{trans} we
get
\be
m^u=\left(\ba{c} 0\\ m\ea\right) \to m'^u=
\left(\ba{c} -m\\ 0\ea\right),~~
{\cal{M}}=\left(\ba{cc}H^{-1\ov2} &0\\ 0&H^{1\ov2}\ea\right)\to 
{{\cal M}'}=\left(\ba{cc}H^{1\ov2} &0\\ 0&H^{-1\ov2}\ea\right),
\lll{dual1}
\ee
while eight-dimensional metric and the dilaton  remain invariant. The
transformed mass vector $m'^u$ implies that the new
configuration is a solution of a massless IIA
compactified on $T^2$   
with 2-form flux along $T^2$.

Now lifting the  solution \eqn{dual1} back  to ten dimensions, 
we get the following type IIA
configuration (we write them  with a prime)
\bea\label{news}
&&d \hat s'^2= H^{-1/2} (-dt^2 +
\sum_{i=1}^6dx_i^2)  + H^{1/2}(dz^2 + dy_1^2+dy_2^2 )\ ,
\br &&2\hat\f'=-{3\ov2} \ln  H\ ,\qquad \hat F'_{(2)}={m}~ dy_1\wedge
dy_2\ .
\eea
According to our ansatz in \eqn{ansatzz} it corresponds to $\hat 
A'_{(1)}=m~ y_{[1}dy_{2]}$. Since under the $SL(2,R)$ transformation 
\eqn{dual1}, $G\to1/G$, 
it amounts to making T-duality along both the torus directions,
therefore
the background  \eqn{news} represents an stack of D6-branes in ten
dimensions
filling the transverse $T^2$ and have
non-trivial 2-form flux. \footnote{Note, 
there is not much distinction between these
  domain-wall-type D6-branes and the usual (asymptotically flat)
 D6-branes which depend on all
  the transverse coordinates and are magnetic duals of D0-branes. The
  domain-wall-type D6-branes are equally possible in type 
IIA theory and are BPS objects.}
 Since this solution is obtained by incorporating T-duality
transformation \eqn{trans} the number of preserved
supercharges remains unchanged. Soon we will show
 that the type IIA solution in 
\eqn{news} becomes an instanton in M-theory set-up.

Thus by making an $SL(2,R)$  transformation
we have transformed 8-brane solution of m-IIA 
into a 6-brane solution of ordinary type IIA
which is supported by a non-trivial 2-form flux.
Thus, the  8-dimensional perturbative duality
interpolates between vacua of massive type IIA  
and ordinary type IIA. It  is 
similar to the situation encountered 
in the case of massive type II duality in
$D=9$ \cite{berg1}. 
Further solutions in $D=8$ and $D=10$ can be generated 
by using other elements of the duality group 
which mix the mass $m$ with the flux $m_{12}$. We shall consider one
example
of this type in the next section.
\vskip.5cm 

\noindent{\it\large M-theory instanton:}
Since ordinary type IIA theory compactified over 
$T^2$ is equivalent to M-theory on 
$S^1\times T^2$,  the solutions of
massive IIA can be lifted to eleven dimensions by first mapping
them to the solutions of  ordinary IIA by using the $SL(2,R)$ element 
\eqn{U3} and then lifting them to 
eleven dimensions. The configuration in 
eq.\eqn{news} is a massless IIA
background and  can be lifted to eleven dimensions.
Correspondingly we get the following eleven dimensional solution 
\bea\label{M5}
ds_{11}^2&=&e^{4\hat\f\over3}(dx_{11}+
2 \hat A_Mdx^M)^2+e^{-{2\hat\f\over3}}ds_{10}^2  \br
&=&  H^{-1}\left(dx_{11}+m(y_1dy_2-y_2dy_1) \right)^2 
+H(dz^2+dy_1^2+dy_2^2)- dt^2 +
\sum_{i=1}^6dx_i^2 \ ,\br
&& H=const.+2m|z-z_0|\ ,
\eea
where $x_{11}$ is the coordinate of 11-dimensional circle $S^1$, $y_1$
and 
$y_2$ are also periodic while other 
fields vanish. This is a pure gravity 
solution in eleven dimensions and has the geometry which is a product 
of a  4-dimensional euclidian instanton (we clarify next why we call
it an instanton), ${\cal E}^4$, and a 
7-dimensional  Minkowski space, $M_7$. 

Let us focus on the properties of the instanton line element in the
above, 
we rewrite it as   
\bea
&&ds^2_{Instanton}=  H^{-1}(d\tau+ K)^2
+H(dz^2+dy_1^2+dy_2^2),\br
&& H=const. + 2m |z-z_0|,~~
K= m(y_1dy_2-y_2dy_1)
\lll{inst}
\eea
with $\tau=x_{11}$ being periodic. 
At closer examination we find that this   ($y_1,~y_2$ are
compact coordinates but their radii can be taken large enough)
looks like a domain-wall-type
generalization of Taub-NUT instantons \cite{gh} 
\be
ds^2_{T-N}=  V^{-1}(d\tau+\Omega(x))^2
+V(dx_1^2+dx_2^2+dx_3^2),
\lll{inst1}
\ee
in which
$V(x)=\ep+{2m\ov|x-x_0|}$, with $V$ and the 1-form $\Omega$  satisfying
$ dV=~^\ast d\Omega\ $. 
The Hodge-dual is defined over flat transverse $x$-space.
Precisely in the same way,  $H$ and $K$  in \eqn{inst} also satisfy an 
identical relation 
 $$ dH=~^\ast d K\ ,$$
where Hodge-dual is defined over flat $z,~y_1,~y_2$ space.
Moreover the spin connections and the 
curvature 2-forms of \eqn{inst} are 
self-dual 
\bea
&&\o^1~_2=\o^0~_3=-{m\ov H^{3/2}}e^0,~~
\o^3~_1=\o^0~_2=-{m\ov H^{3/2}}e^1,~~ 
\o^2~_3=\o^0~_1={m\ov H^{3/2}}e^2\ , \br
&&R^1~_2=R^0~_3={4m^2\ov H^3}(e^3~e^0+e^2~e^1),~~
R^3~_1=R^0~_2={2m^2\ov H^3}(e^3~e^1+e^0~e^2),~~\br && 
R^2~_3=R^0~_1={2m^2\ov H^3}(e^2~e^3+e^0~e^1) \ ,
\eea
with the basis $e^0=H^{-1\ov2} (d\tau+ my_1dy_2-my_2dy_1),~e^1=
H^{1\ov2} dy_1,~e^2=H^{1\ov2} dy_2,~ e^3=H^{1\ov2} dz$.
Thus the line element in \eqn{inst} is essentially a Taub-NUT
instanton in four dimensions, and is of a new kind in that it has a
domain-wall
type extent. Thus we  call it a 
{\it domain-wall-instanton}. 

We note that a 11-dimensional solution almost similar to  
\eqn{M5} also 
appear in  \cite{cow,lavpop}, perhaps authors  miss in 
identifying them as  instanton line
elements. The line element in \eqn{M5} differs  in the structure of 
the connection 1-form $K$ from the previous occasions. The present
form of $K$ crucially
depends on our reduction ansatz in \eqn{ansatzz}. 
This affects the periodicity of the
coordinate, $\tau$, of the circle $S^1$ which is fibred over the base 
$z,~y_1,~y_2$. 
We note that the periodicity of  coordinate $\tau$ is 
independent of the periodicity of the $T^2$ coordinates $y_1$ and $y_2$ 
which is not the case discussed in \cite{lavpop}. That is 
we can choose the periodicity of  $T^2$ independent  of $S^1$. It is 
essential in order to make the connection between m-IIA and M-theory.
The duality element \eqn{U3} takes  $T^2$ to its 
inverse size when we relate D8-brane with D6-brane which is  
subsequently oxidised to M-theory. Thus to get a decompactified D8-brane 
of m-IIA,  $T^2$ on 
the M-theory side must be taken to zero size  independently of 
the size 
of $S^1$ (which also goes to zero  in type IIA limit), 
and the vice-versa. This  zero area limit has also been 
imphasized in the work of Chris Hull \cite{hull}
 although from the perspective of F-theory, as the analysis 
there  deals 
 with generalized compactification of IIB on $S^1$ \cite{berg1}.  While
in 
present set-up we 
do not encounter that situation as we  all the time remain in IIA
set-up.     

Thus D8-branes of massive type IIA theory emerge from purely geometrical 
considerations of the 
M-theory such that it involves  `domain-wall-instanton'. 
Thus we are tempted to claim that
 M-theory compactifications on  $M_7\times 
{\cal E}_4$ will correspond to massive IIA  compactification on 
2-torus.  
The sketch below is an effort to summarize the whole picture

\vskip1cm

\begin{tabular}{rcc}
~&~~~ & \fbox{\large\rm M-theory~ on~ $M_7\times{\cal E}_4$ }\\
~&~~~& ~~~~~~~~~$\updownarrow$ ~\\    
\fbox{\large\rm massive\ IIA\ on~ $T^2$} & $\stackrel{(G\leftrightarrow 
{1\over  G})}
{\longleftrightarrow} $
&\fbox{\large\rm \ IIA~ on~ $T^2$ $\&$ 2-form-Flux}
\end{tabular}
\vskip1cm

\section{D6-D8 bound state}
In the above we have studied the background configurations 
which  either have mass or flux but not the both.  
By making a more general $SL(2,R)$ transformation we can  generate
vacuas
in which both mass and the flux are nontrivial. Let us make the
following
$SL(2,R)$ rotaion 
\be
\label{U4}
\L=\left(\ba{cc} cos\t&~sin\t\\ -sin\t&~cos\t\ea\right)\ .
\ee
Using the transformations \eqn{trans} and applying them on the 
configuration in \eqn{sixs}, we
get the transformed configuration with
\be
 m^u=
\left(\ba{c} m~sin\t\\ m~cos\t\ea\right),~~
{{\cal M}}=\left(\ba{cc}H^{-1\ov2}cos^2\t+H^{1\ov2}sin^2\t
  &~~(-H^{-1\ov2}+H^{1\ov2})cos\t~sin\t  \\ 
(-H^{-1\ov2}+H^{1\ov2})cos\t~sin\t
 &~~H^{-1\ov2}sin^2\t+H^{1\ov2}cos^2\t \ea\right),
\lll{dual2}
\ee
while the eight-dimensional metric and dilaton remain same as in
\eqn{sixs}.
Uplifting \eqn{dual2} to ten dimensions gives us
\bea\label{news1}
&&d \hat s^2_{10}= H^{1/2}\left\{ H^{-1} (-dt^2 +
dx_i^2)  + H'^{-1}( dy_1^2+dy_2^2 )+dz^2\right\}\ ,
\br &&e^{2\hat\f}=g_s^2 H^{-3/2} H'^{-1}\ ,\qquad \hat 
F_{(2)}=-{1\ov{g_s}}m~sin\t H'^{-1}
dy_1\wedge dy_2\ ,\br
&& 2 \hat B_{{y_1}{y_2}}= tan\t (1+{1\ov H-1}{1\ov cos^2\t})^{-1}, 
\eea
where  $H=const + 2m |z-z_0|$ 
and $H'=cos^2\t~(H-1)+1$. We have introduced the
parameter $g_s$ which represents string coupling constant. Since this
configuration is a solution of massive IIA theory (with new mass 
parameter ${m~cos\t \over g_s}$) and also has nontrivial
2-form flux,  our interpretation is that \eqn{news1}
 represents a bound state of D6 and D8
branes. Moreover it preserves 16 supercharges. Note that the asymptotic 
($z\to\infty$) value of $\hat B$ is  proportional to $tan\t$ which 
is  usually the case 
with D$(p-2)$-D$p$  bound states for $(2\le p\le 6) $
\cite{roy}.  It would be therefore
interesting to study the non-commutative Yang-Mills (NCYM) decoupling
limit 
\cite{witten} for the D6-D8 bound state in \eqn{news1}, which we leave
for 
a later investigation \cite{singh}.

\section{Compactification on $T^4$} 

Through a similar procedure a generalized compactification 
of m-IIA theory with R-R fluxes can 
be carried out on a four-torus as well. On a four-torus we have six 
2-cycles  and one 
4-cycle. So we can switch on  fluxes correspoding to each of these 
cycles. This will add seven new flux parameters in the six-dimensional
theory. 
These 
seven parameters and the mass will combine into an eight-dimensional 
spinorial  representation 
of the $SO(4,4)$ T-duality group. A generalized Kaluza-Klein ansatz can
be
written as
\bea
&& \hat A(x,y)=  A(x)+ (a_m(x)-{1\ov2}m_{mn}y^n) dy^m  ,\br
&&\hat C(x,y)= C_{(3)}(x)+ 
T_{(2)m}(x) dy^m+{1\ov2} V_{(1)mn}(x) dy^mdy^n \br
&&\hskip1cm+ {1\over3!}(s_{(0)mnp}(x) 
-{1\ov4}\beta_{mnpq} y^q)dy^mdy^ndy^p~,
\lll{ansatz}
\eea
where all the scalar-forms are allowed to retain up to a linear 
dependence on the coordinates of the torus. Flux parameters $m_{mn}$ and 
$\beta_{mnpq}$ are completely 
antisymmetric in their indices, $m,n=1,2,3,4$.  
Then
\bea
&& \hat d\hat A= \cd A + \cd a_m \eta^m +{1\ov 2}m_{mn} \e^m\e^n,\br
&&\hat d\hat C=\cd C_{(3)}(x)+\cd T_{(2)m} \eta^m +{1\ov2} 
\cd V_{(1)mn}\e^m\eta^n \br&&\hskip1cm
+{1\ov3!}\cd s_{(0)mnp}\e^m\e^n\e^p+
{1\ov 4!}\beta_{mnpq}\e^m\e^n\e^p\e^q 
\lll{ansatz1b}
\eea
where various derivatives $\cd$ are given by
\bea
&& \cd A= dA -da_m K^m +{1\ov2}m_{mn}K^mK^n,~~ \cd a_m=da_m + m_{mn}
K^n, \br
&& \cd s_{mnp }= ds_{mnp}  +\beta_{mnpq}K^q \br
&& \cd V_{mn }= dV_{mn} -d s_{mnp} K^p +{1\ov 2}
\beta_{mnpq}K^pK^q, \br
&& \cd T_{m }= dT_m +d V_{mn} K^n +{1\ov 2}ds_{mnp}K^nK^p+{1\ov 
3!}\beta_{mnpq}K^nK^pK^q, \br
&& \cd C= dC -dT_m K^m+{1\ov 2!} dV_{mn}K^m K^n
-{1\ov 3!}ds_{mnp}K^mK^nK^p  +{1\ov 
4!}\beta_{mnpq}K^mK^nK^pK^q~.\br
&&
\eea
Thus we see explicitely that even if the potential depends upon the 
compact torus 
coordinates but this dependence is dropped out in their derivatives so
also
in the field equations.  
Correspondingly, massive field strengths in \eqn{a22} become
\bea
&&\hat F_{(2)}=(\cd A+2m B) +(\cd a_n+2m\bar A_m)\e^m +{1\over2} 
(m_{mn}+2mB_{mn})\e^m\e^n~, \br 
&&\hat F_{(4)}=(\cd C+ 2B\cd A+2m BB) +(\cd T_m+ 2B\cd a_m+ 2 A_m\cd
A+4mB 
\bar 
A_m )\e^m \br 
&&\hskip1.5cm +{1\over2} 
(\cd V_{mn} -4\bar A_m\cd a_n +2B_{mn}\cd A +2Bm_{mn} 
+4 mBB_{mn}-4m\bar A_m \bar A_n)\e^m\e^n~ \br 
&&\hskip1.5cm
+{1\ov3!} (\cd s_{mnp} +6B_{mn}\cd a_p +6 \bar A_m m_{np}+12 m\bar A_m 
B_{np})\e^m\e^n\e^p \br
&&
\hskip2cm +{1\ov4!}(\beta_{mnpq}+12m_{mn}B_{pq}+12m 
B_{mn}B_{pq})\e^m\e^n\e^p\e^q .
\lll{as2}
\eea

As in the case of ordinary type IIA compactified on $T^4$ \cite{sen},
the
8 vector fields $(\bar A_m,~K^m)$ 
coming from the fields in NS-NS sector transform in the vectorial 
representation of $SO(4,4)$, 8 
scalars $(a_m,~s_{mnp})$ transform in the eight-dimensional spinorial 
representation $R_s$, 8 vectors $(A,~ V_{mn},~{\rm dual~of~}C_{(3)})$ 
coming from  the R-R 
sector transform in 
the another spinorial representation $R_c$, 4 tensor fields $T_m$  
split into 
eight (anti)self-dual 3-form field strengths 
and there field equations transform 
covariantly. Finally from the examination of various field 
strengths we determine that eight masses and fluxes 
$(m,~m_{mn},~\beta_{mnpq})$  fit in the eight dimensional 
representation $R_c$. To see this  explicitely let us 
write down the contribution from purely mass terms which follow 
from the Lorentz scalar contractions of the last terms of the two 
equations in \eqn{as2},
\bea
&&
 \int (-{1\over2}\hat F_2 ~^{\hat\ast}\hat F_2 -{1\over2}\hat F_4
~^{\hat\ast}\hat F_4- {m~^{\hat\ast} m\ov 2 })\br
&&~~~~~=-{1\ov2} \int d^6x~  e \sqrt{G}\bigg[(m)^2
+{1\ov2!}(m_{mn}+2mB_{mn})(m_{pq}+2mB_{pq})G^{mp}G^{nq}
\br &&~~~~~~~~+{1\ov4!} (\beta_{mnpq}+12m_{[mn}B_{pq]}+12m 
B_{[mn}B_{pq]})^2\bigg] +~{\rm other~ terms}
\lll{cosmo}
\eea
where internal indices are contracted with the metric $G_{mn}$ on the 
four-torus.
Thus the mass $m$ and seven fluxes arrange themselves into a 
representation, $R_c$, 
same as the eight RR vector potentials, and transform under 
$SO(4,4)$ accordingly. The terms on the right hand side of
eq.\eqn{cosmo} 
 represent $SO(4,4)$ invariant contribution
of the mass (or cosmological constant) terms to six-dimensional massive
II 
theory. 
In other words  the massive 
six-dimensional type II theory obtained in this way
will have $SO(4,4)$ invariance
 provided the fluxes and masses
transform in the 8 dimensional spinorial representation of $SO(4,4)$.    

To recall, situation here is analogous to the case of massive type IIA
theory
compactified on $K_3$ \cite{hls} where mass and the RR fluxes fit into a
vectorial representation of $SO(4,20)$ duality group. However the number
of
supercharges are double here. The duality group $SO(4,4)$ mixes mass and
various fluxes
and all of these fluxes on $T^4$  
can be lifted to eleven
dimensions, case by case. In the picture below we have summarized a
case  
where 4-form flux on $T^4$ is lifted to eleven dimensions. The 4-form
flux
gives rise to 4-form flux in eleven dimensions along the compact four
torus. 
The  vacuum solutions with  2-form fluxes would be interesting to study 
as they will give rise to pure gravity
configurations in eleven-dimensions like the domain-wall-instanton 
in the previous case of $T^2$.

\vskip1cm

\begin{tabular}{rcc}
~&~~~ & \fbox{\large\rm M-theory~ on~ $S^1\times T^4$ $\&$ 
4-form-Flux}\\
~&~~~& ~~~~~~~~~$\updownarrow$ ~\\    
\fbox{\large\rm massive\ IIA\ on~ $T^4$} & $\stackrel{(G\leftrightarrow 
{1\over G})}
{\longleftrightarrow}$
&\fbox{\large\rm \ IIA~ on~ $T^4$ $\&$ 4-form-Flux}
\end{tabular}
\vskip1cm

Above sketch is quite  similar to a sketch in \cite{hls} where M-theory 
compactification on  $S^1\times K3$ with flux is discussed.  
The M-theory solution for  $S^1\times T^4$ compactification with 4-form 
flux can be obtained by  replacing $K3$ line element by a 
 $T^4$ line element in the equations (4.2) of \cite{hls}.  

\section {Summary}
In this paper we have studied the  
$T^2$ compactification of 
ten-dimensional massive type IIA theory 
with  Ramond-Ramond background flux corresponding to 2-form field
strength. 
We found that the resulting eight-dimensional theory is a $SL(2,R)\times 
SL(2,R)$ symmetric 
massive supergravity theory. 
The mass and the flux parameters transform under $SL(2,R)$ accordingly.
Thus the perturbative T-duality survives even at
the massive level when appropriate masses and fluxes are switched
on. 
Next we have shown that this duality symmetry
interpolates between vacua of 
massive type IIA compactified on $T^2$ and vacua of ordinary type IIA  
compactified on $T^2$ with a RR 2-form flux. The 
wrapped D8-brane solution of massive
type IIA turns out to be T-dual
to D6-brane solution of ordinary type IIA theory.
This relationship between massless and massive IIA vacuas on $T^2$ also 
suggests an 11-dimensional interpretation of massive IIA theory. We find
that
the D8-brane is related to pure gravity vacua of M-theory which is a
direct product of seven-dimensional Minkowski space and 4-dimensional
instaton. This Ricci-flat instanton turns out to be a domain-wall
generalization of Gibbons-Hawking
multi-center Taub-NUT instanton. Thus the compactification of M-theory
on such
instantons will tell us more deeply about the spectrum of massive type
II sugras in
lower dimensions. 

We have shown that D6-D8 brane bound states can also 
exist in massive type IIA similar to the ordinary
  D$(p-2)$-D$p$ brane bound states 
 in constant magnetic $B$-field background. It would be interesting to 
investigate if there are corresponding  non-commutative Yang-Mills 
theories \cite{witten} 
for D6-D8  bound state. 

We also  discuss the compactification of massine type IIA on $T^4$.
We find that the mass and the seven R-R fluxes organize themselves into
an 
eight-dimensional  representation of duality group
$SO(4,4)$. Many of the features of these fluxes will be 
similar to the case of $T^2$ compactifications.

Finally let us say something about strong-weak dualities. The
full duality symmetry group of type II theory in eight dimensions is
$SL(2,Z)\times SL(3,Z)$ which includes the strong-weak duality elements.
This non-perturbative duality group cannot be restored in our set-up as 
we cannot generate more flux parameters than what we already have.
However, there is an $SL(3,R)$ symmetric massive type II action worked
out in
\cite{amees} from  $T^3$ compactification of a massive 11-dimensional
sugra \cite{bs}. 
\vskip1cm
\noindent{\bf Note added:} After this work was communicated I came 
to know about the overlap between  11-dimensional solution in 
section~4 and the works in \cite{ 
gibbons,pope}.

\acknowledgments
I would like to thank Jan Louis, Thomas Mohaupt and Marco Zagermann 
for many useful discussions. I am also  grateful to M. Zagermann
for a very careful reading of the manuscript.  
This work is supported by AvH (the Alexander von Humboldt foundation).

\end{document}